\documentclass[fleqn,usenatbib]{mnras_tex_edited}

\usepackage{amssymb,amsmath,verbatim,mathtools,needspace,enumitem,etoolbox,graphicx,physics,microtype,natbib,url,hyperref,mathrsfs,bm,ulem,lipsum}
\usepackage{orcidlink}

\newcommand{\ssim}{\mathchar"5218\relax\,}

\renewcommand{\emph}[1]{\textit{#1}}

\newcommand{\approptoinn}[2]{\mathrel{\vcenter{
  \offinterlineskip\halign{\hfil$##$\cr
    #1\propto\cr\noalign{\kern2pt}#1\sim\cr\noalign{\kern-2pt}}}}}

\renewcommand{\vec}[1]{\boldsymbol{#1}}
\newcommand{\vlambda}{\vec{\lambda}}
\newcommand{\vtheta}{\vec{\theta}}
\newcommand{\vd}{\vec{d}}
\newcommand{\Nobs}{N_\mathrm{obs}}
\newcommand{\Pdet}{P_\mathrm{det}}
\newcommand{\fR}{f_\mathrm{R}}
\newcommand{\fS}{f_\mathrm{S}}

\def\apj{{Astrophys.~J.}}\def\apjl{{Astrophys.~J.~Lett.}}
\def\apjs{{Astrophys.~J.~Supp.}}
\def\ao{{Appl.~Opt.}}

\def\aap{{Astron.~Astrophys.}}

\def\mnras{{Mon.~Not.~R.~Astron.~Soc.}}

\def\prd{{Phys.~Rev.~D}}

\def\prx{{Phys.~Rev.~X}}
\def\prl{{Phys.~Rev.~Lett.}}
\def\pasa{{PASA}}

\def\physrep{{Phys.~Rep.}}

\def\cqg{{Classical~Quant.~Grav.}}
\def\natastro{{Nat. Astron.}}

\newcommand{\milan}{Dipartimento di Fisica ``G. Occhialini'', Universit\'a degli Studi di Milano-Bicocca, Piazza della Scienza 3, 20126 Milano, Italy}
\newcommand{\infn}{INFN, Sezione di Milano-Bicocca, Piazza della Scienza 3, 20126 Milano, Italy}
\newcommand{\bham}{School of Physics and Astronomy \& Institute for Gravitational Wave Astronomy, University of Birmingham,\vspace{-0.05cm}\\$\;$Birmingham, B15 2TT, UK}
\newcommand{\harvard}{Center for Astrophysics \textbar{} Harvard \& Smithsonian, 60 Garden St., Cambridge, MA 02138, USA}

\title[Which black hole formed first?]{Which black hole formed first? Mass-ratio reversal in massive binary stars from gravitational-wave data}

\author[Mould et al.]{Matthew Mould\thanks{\href{mailto:mmould@star.sr.bham.ac.uk}{mmould@star.sr.bham.ac.uk}}\,\orcidlink{0000-0001-5460-2910}$^{1}$,
Davide Gerosa\,\orcidlink{0000-0002-0933-3579}$^{2,3,1}$,
Floor S. Broekgaarden\,\orcidlink{0000-0002-4421-4962}$^{4}$,
\newauthor
Nathan Steinle\,\orcidlink{0000-0003-0658-402X}$^{1}$
\medskip
\\
$^{1}$\bham\\
$^{2}$\milan\\
$^{3}$\infn\\
$^{4}$\harvard
}

\pubyear{\the\year{}}

\begin{document}
\label{firstpage}
\pagerange{\pageref{firstpage}--\pageref{lastpage}}
\maketitle

\begin{abstract}

Population inference of gravitational-wave catalogues is a useful tool to translate observations of black-hole mergers into constraints on compact-binary formation. Different formation channels predict identifiable signatures in the astrophysical distributions of source parameters, such as masses and spins. One example within the scenario of isolated binary evolution is mass-ratio reversal: even assuming efficient core--envelope coupling in massive stars and tidal spin-up of the stellar companion by the first-born black hole, a compact binary with a lighter, non-spinning first-born black hole and a heavier, spinning second-born black hole can still form through mass transfer from the initially more to less massive progenitor. Using current LIGO/Virgo observations, we measure the fraction of sources in the underlying population with this mass--spin combination and interpret it as a constraint on the occurrence of mass-ratio reversal in massive binary stars. We modify commonly used population models by including negligible-spin subpopulations and, most crucially, non-identical component spin distributions. We do not find evidence for subpopulations of black holes with negligible spins and measure the fraction of massive binary stars undergoing mass-ratio reversal to be consistent with zero and $<32\%$ ($99\%$ confidence). The dimensionless spin peaks around $0.2$--$0.3$ appear robust, however, and are yet to be explained by progenitor formation scenarios.
\end{abstract}


\section{Introduction}
\label{sec: introduction}

Gravitational-wave (GW) population studies have become a common approach to learn about the astrophysical distribution of merging stellar-mass binary black holes (BHs). Using catalogues of detected GW signals~\citep{2019PhRvX...9c1040A,2021arXiv210801045T,2021arXiv211103606T,2021PhRvX..11b1053A} and hierarchical Bayesian inference~\citep{2019MNRAS.486.1086M,2020arXiv200705579V}, the parameters of a given population model can be constrained. Many independent analyses have used parametric (e.g., \citealt{2021arXiv211103634T}), semiparametic~\citep{2022PhRvD.105l3014S,2022ApJ...924..101E,2022MNRAS.509.5454R,2022ApJ...928..155T}, and simulation-based modelling~\citep{2018PhRvD..98h3017T,2020PhRvD.101l3005W,2021ApJ...910..152Z,2022arXiv220303651M} to measure the distributions of source masses, spins, redshifts, and correlations between parameters~\citep{2021ApJ...922L...5C,2021ApJ...912...98F,2022A&A...665A..59B,2022ApJ...932L..19B}.

In some cases, different approaches have led to contradictory results. \cite{2021arXiv211103634T,2021ApJ...913L...7A} infer that BHs have non-zero spins with support for large misalignments with respect to the binary orbital angular momentum, while \cite{2021ApJ...921L..15G} and \cite{2021PhRvD.104h3010R} report a large population of non-spinning BHs or a lack of evidence for misaligned spins. Such differences become important when attempting to relate the results of GW inference with predictions of compact binary formation, in which distinct channels leave identifiable imprints on the population of detectable mergers~\citep{2022PhR...955....1M,2021hgwa.bookE...4M}.

In the classical picture of isolated stellar binary evolution, the initially more massive star evolves more rapidly and thus collapses first to form the heavier BH. If angular momentum is lost from the outer envelope of the massive star after efficient transport from the stellar core~\citep{2002A&A...381..923S,2019MNRAS.485.3661F,2020A&A...636A.104B} the heavier BH is expected to have negligible spin~\citep{2018A&A...616A..28Q,2019ApJ...881L...1F}. Before the second BH forms, its progenitor may be spun up due to tidal torques from the first-born BH~\citep{2016MNRAS.462..844K,2017ApJ...842..111H,2018A&A...616A..28Q,2018MNRAS.473.4174Z,2020A&A...635A..97B,2020A&A...636A.104B}. The less (more) massive BH in the resulting binary therefore has significant (negligible) spin.

On the other hand, if there is significant mass transfer from the initially more to less massive star, the binary mass ratio can be ``reversed'' and the first (second) collapse of the now less (more) massive star can form the lighter (heavier) BH. This mechanism has frequently been suggested in the literature to explain Algol binaries and compact object binaries with young  neutron stars (e.g., \citealt{1996A&A...309..179P,2000ApJ...543..321K,2000A&A...355..236T, 2004MNRAS.354L..49S,2018A&A...619A..53T,2020Sci...367..577V}). Crucially, this scenario of mass ratio reversal (MRR) leads to GW detections of binary BHs with the more (less) massive BH having non-negligible (negligible) spin~\citep{Broekgaarden_2022,2022ApJ...933...86Z}, as well as different spin misalignments \citep{2013PhRvD..87j4028G}. While the individual BH spins are typically poorly measured~\citep{2021arXiv211103606T}, in principle the occurrence of MRR can be constrained using GW observations.

To this end, we present a suite of GW population analyses with models designed to identify via spin measurements binary BHs that have undergone MRR. The key ingredient is that we must no longer assume the two BH spins are identically distributed as is done in current state-of-the-art analyses \citep{2021arXiv211103634T}. We consider minimal extensions of widely used spin models~\citep{2017PhRvD..96b3012T,2019PhRvD.100d3012W,2021arXiv211103634T}, accounting also for the possibility of small or non-spinning components~\citep{2021ApJ...921L..15G}. From the underlying population we measure the fraction of sources in which the heavier BH has larger spin and, by thresholding the spin magnitudes or including a peak at zero spins, the fraction in which either one or both components have negligible spin. These measurements give insight in the fractions of binary BHs undergoing MRR.

In Sec.~\ref{sec: methods}, we summarize our modelling and inference procedure. In Sec.~\ref{sec: results}, we present the results of our analyses, in particular the fractions of sources with negligible spins and the MRR fraction. We discuss our findings and their caveats in Sec.~\ref{sec: discussion}.

\section{Methods}
\label{sec: methods}

\subsection{Population inference}
\label{sec: population inference}

We model GW events as independent draws from an inhomogeneous Poisson process subject to detection biases. For the $\Nobs=69$ events with false alarm rates (FARs) $<1~\mathrm{yr}^{-1}$ and source parameters $\vtheta_i$ as measured from strain data $\vd=\{\vd_i\}$ ($i=1,...,\Nobs$) in the first (O1), second (O2), and third (O3) observing runs, the hierarchical Bayesian posterior distribution of population parameters $\vlambda$ characterizing a given astrophysical model $p(\vtheta|\vlambda)$ is given by $p(\vlambda|\vd) = \pi(\vlambda) \mathcal{L}(\vd|\vlambda) / \mathcal{Z}(\vd)$. We take uninformative priors $\pi(\vlambda)$ (see Sec.~\ref{sec: spin models}). The evidence for the GW data \emph{given the population model} is $\mathcal{Z}(\vd) = \int \pi(\vlambda) \mathcal{L}(\vd|\vlambda) \dd\vlambda$. The population likelihood is given by~\citep{2019MNRAS.486.1086M,2020arXiv200705579V}
\begin{align}
\mathcal{L}(\vd|\vlambda) \propto \prod_{i=1}^{\Nobs} \frac{\int \mathcal{L}(\vd_i|\vtheta_i) p(\vtheta_i|\vlambda) \dd\vtheta_i}{\int \Pdet(\vtheta) p(\vtheta|\vlambda) \dd\vtheta}
\, ,
\label{eq: likelihood}
\end{align}
where each $\mathcal{L}(\vd_i|\vtheta_i)$ is a single-source likelihood and $\Pdet(\vtheta)$ is the detection probability of a source given its parameters $\vtheta$. The posterior has been marginalized over the merger rate normalization with a log-uniform prior~\citep{2018ApJ...863L..41F}.

We evaluate Eq.~(\ref{eq: likelihood}) using Monte Carlo integrals. We rewrite the likelihoods as $\mathcal{L}(\vd_i|\vtheta_i) = \mathcal{Z}(\vd_i) p(\vtheta_i|\vd_i) / \pi(\vtheta_i)$ and use public LIGO/Virgo data to compute the numerator as a summation over samples $\vtheta_i$ drawn from the single-event posteriors $p(\vtheta_i|\vd_i)$.\footnote{Samples used in the \textsc{Power Law + Peak} analysis from \href{https://zenodo.org/record/5655785}{zenodo.org/record/5655785}.} The denominator is computed by reweighting to our population models the recovered sources, for which $\Pdet(\vtheta)=1$, from an injected set of signals.\footnote{Dataset \texttt{o1+o2+o3\_bbhpop\_real+semianalytic-LIGO-T2100377-v2} from \href{https://zenodo.org/record/5636816}{zenodo.org/record/5636816}.} For events detected in O3 the threshold for recovery is set to $\mathrm{FAR}<1~\mathrm{yr}^{-1}$ in at least one detection pipeline, while for events from O1 and O2 we require a network signal-to-noise ratio (SNR) $>10$~\citep{2021arXiv211103634T}. We account for uncertainty in both integrals due to finite sampling as in \cite{2021arXiv211103634T} (see also \citealt{2019RNAAS...3...66F,2022ApJ...926...79G}).

The source-frame primary BH masses and mass ratios are modelled using the \textsc{power law + peak} model, while the model for redshifts $z$ corresponds to a merger rate density per unit comoving volume and source-frame time that is a power law in $1+z$. The parameters and their prior distributions are as in \cite{2021arXiv211103634T}. Our inference on the mass and redshift parameters is unchanged with respect to any of the following spin models. The population-level posterior is sampled using \textsc{gwpopulation}~\citep{2019PhRvD.100d3030T}, \textsc{bilby}~\citep{2019ApJS..241...27A}, and \textsc{dynesty}~\citep{2020MNRAS.493.3132S}.

\subsection{Spin models}
\label{sec: spin models}

\subsubsection{Identical spins}
\label{sec: identical spins}

The \textsc{default} spin parametrization used by \cite{2021arXiv211103634T} models the dimensionless spin magnitudes $\chi_{1,2}$ (corresponding to BH masses $m_1\geq m_2$) as independent and identical beta distributions that have mean $\mu$ and variance $\sigma^2$, with uniform priors on $[0,1]$ and $[0.005,0.25]$, respectively. Additional prior constraints on the standard beta distribution shape parameters $\alpha,\beta>1$ enforce regularity at the prior boundaries (one could alternatively sample these parameters directly as in, e.g., \citealt{2022arXiv220400968V}). The two polar spin misalignments are modelled independently of the magnitudes \emph{but not of each other}, being distributed in cosines as a mixture between isotropic (flat in cosine) and preferentially aligned (unit-mean Gaussian truncated on $[-1, 1]$) components, indicative of isolated and dynamical formation, respectively (e.g., \citealt{2010CQGra..27k4007M}). The prior on the mixing fraction $\zeta$ is uniform in $[0,1]$ while that of the aligned-component standard deviation $\tau$ is uniform in $[0.1,4]$.

\subsubsection{Non-identical spins}
\label{sec: non-identical spins}

To asses the difference between BH spin magnitudes, the first extension we consider is to relax the assumption of identical distributions. In this \textsc{nonidentical} model the spin magnitudes employ the same independent beta parametrization but are no longer identical, with means $\mu_i$ and variances $\sigma_i^2$ ($i=1,2$). We additionally allow non-identical alignments $\tau_i$; however, since spin magnitudes and tilts are independent, this will not directly affect inferences on $\chi_{1,2}$. The mixing fraction $\zeta$ between aligned and isotropic tilts is kept identical for both BHs since they must have formed in the same environment. We impose priors as above.

\subsubsection{Zero-spin peaks}
\label{sec: zero-spin peaks}

Since predictions of isolated binary stellar evolution include BHs with small spins as discussed in Sec.~\ref{sec: introduction}, we also consider a model that explicitly includes such subpopulations. In the spirit of \cite{2021ApJ...921L..15G}, we modify the \textsc{nonidentical} spin model by adding zero-mean Gaussians truncated within $0\leq\chi_{1,2}\leq1$ to the spin magnitude distributions. The prior of the standard deviations $\omega_i$ is uniform with $0.01\leq\omega_i\leq0.05$ and the mixing fractions $f_i$ between the beta distributions and zero-spin peaks are uniform in $[0,1]$. We keep the \textsc{nonidentical} distribution of spin tilts and allow for the possibility of just one (non-) spinning BH by taking the two dimensionless spin magnitude distributions to be independent of each other (such that the joint distribution is a product of two mixtures, rather than a mixture of two products). We refer to this model as \textsc{nonidentical + zeros}.

\section{Results}
\label{sec: results}

Here we focus on the differences between the spin magnitude distributions of the two BHs, as is important in the MRR scenario. We present the relevant posterior distributions of the spin hyperparameters and spin magnitudes for the \textsc{nonidentical} and \textsc{nonidentical + zeros} models in Appendix~\ref{app: spin posteriors}. In the following we quote numerical results with either medians and 90\% symmetric intervals, or the upper 99\% confidence bound in the case of one-sided posteriors.

\subsection{non-identical spins}
\label{sec: results non-identical spins}

\begin{figure}
\includegraphics[width=\columnwidth]{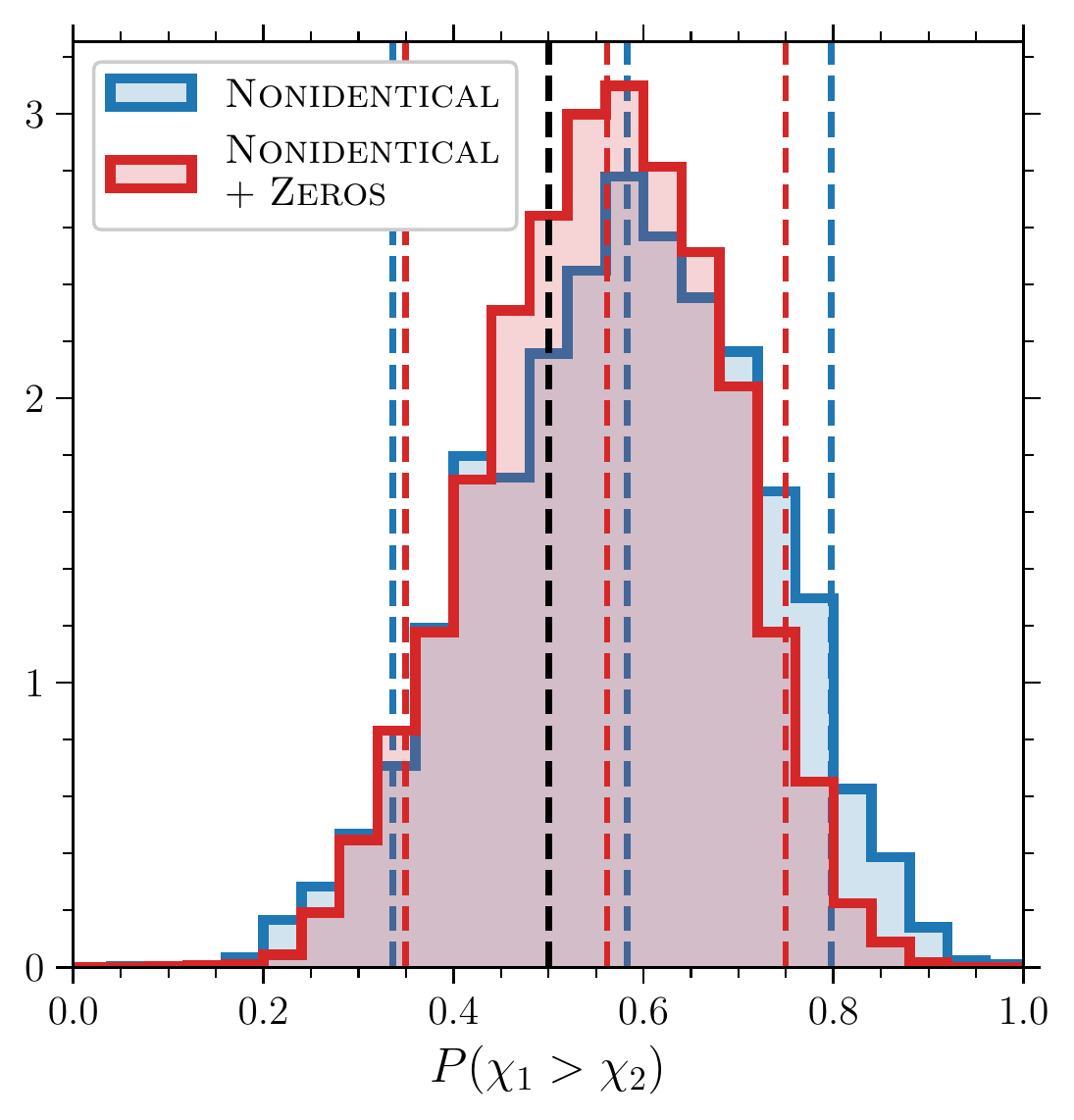}
\caption{The probability that the heavier BH is more rapidly spinning in our \textsc{nonidentical} (blue) and \textsc{nonidentical + zeros} (red) models. The uncertainty is due to the uncertainty in the model hyperparameters, as measured from GW data. The coloured vertical dashed lines correspond to the medians and symmetric 90\% confidence intervals. The vertical dashed black line at $P(\chi_1>\chi_2)=1/2$ corresponds to the result for identical spin distributions.}
\label{fig: 1gtr2}
\end{figure}

We first consider the \textsc{nonidentical} model, which allows us to ask the question: do the more massive BH components spin more rapidly than the less massive components? Since the spin magnitude distributions are independent of each other and not identical, i.e., $p(\chi_1,\chi_2|\vlambda) = p(\chi_1|\vlambda_1) p(\chi_2|\vlambda_2)$, the probability that the more massive BH spins more rapidly than the less massive BH (in terms of the dimensionless Kerr parameter) is given by
\begin{align}
P(\chi_1>\chi_2|\vlambda) = \int_0^1 \dd\chi_1 p(\chi_1|\vlambda_1) \int_0^{\chi_1} \dd\chi_2 p(\chi_2|\vlambda_2)
\, ,
\label{eq: 1gtr2}
\end{align}
where $\vlambda_i=(\mu_i,\sigma_i)$ are the hyperparameters individually characterizing the spin distributions. Our measurement for the probability of more rapidly spinning primaries, given the Bayesian uncertainty $p(\vlambda|\vd)$ in the hyperparameters of the \textsc{nonidentical} model, is given by the blue distribution in Fig.~\ref{fig: 1gtr2}. We find $P(\chi_1>\chi_2|\vlambda) = 0.58_{-0.25}^{+0.21}$. If the two distributions were instead identical, Eq.~(\ref{eq: 1gtr2}) reduces to $P(\chi_1>\chi_2|\vlambda) = P(\chi_1<\chi_2|\vlambda) = 1/2$. We infer a median value above this mid-point and 71\% posterior support for $P(\chi_1>\chi_2|\vlambda)>0.5$, indicating mild evidence that primary BHs spin more rapidly.

While this measurement compares spin components across the entire unit interval, the isolated binary evolution scenario discussed in Sec.~\ref{sec: introduction} predicts small BH spins. To relate spin measurements more closely to these predictions we place a threshold $\chi_0$ which defines a spin magnitude as negligible. The probabilities that a BH spin lies below or above this threshold are given by
\begin{align}
P(\chi_i<\chi_0|\vlambda) = 1 - P(\chi_i>\chi_0|\vlambda) = \int_0^{\chi_0} \dd\chi_i p(\chi_i|\vlambda)
\, .
\end{align}
We find that $4.8_{-4.5}^{+7.0}\%$ ($4.0_{-3.9}^{+11}\%$) of primary (secondary) BHs have negligible spins when imposing a threshold $\chi_0=0.05$~\citep{Broekgaarden_2022}. These results are one-sided and peak at zero, skewing the posterior measurements away from the boundary. Instead, one can consider these as measurements of $P(\chi_i<\chi_0|\vlambda)\approx0$ with one-sided uncertainties $P(\chi_1<\chi_0|\vlambda)<0.15$ and $P(\chi_2<\chi_0|\vlambda)<0.20$.

Since the spin magnitudes are modelled independently, the joint probability that, e.g., the primary BH spin is non-negligible and the secondary is negligible is simply given by the product $P(\chi_1>\chi_0,\chi_2<\chi_0|\vlambda) = P(\chi_1>\chi_0|\vlambda) P(\chi_2<\chi_0|\vlambda)$. If one assumes that the entire binary BH population formed through isolated evolution and that there is no other mechanism by which primary BHs can acquire spin, this probability can be interpreted as the fraction of MRR sources. On the other hand, sources with $\chi_1<\chi_0$ but $\chi_2>\chi_0$ correspond to the standard isolated binary scenario in this interpretation. Again taking $\chi_0=0.05$, we find $<12\%$ ($<10\%$), or $3.8_{-3.7}^{+11}\%$ ($4.6_{-4.3}^{+6.9}\%$), of the underlying population falls into the MRR (standard) spin scenario. The majority of the posterior support lies away from zero spins with $P(\chi_1,\chi_2>\chi_0|\vlambda)=0.90_{-0.10}^{+0.07}$, while the fraction of sources with two negligible spins is at the subpercent level.

\begin{figure}
\includegraphics[width=\columnwidth]{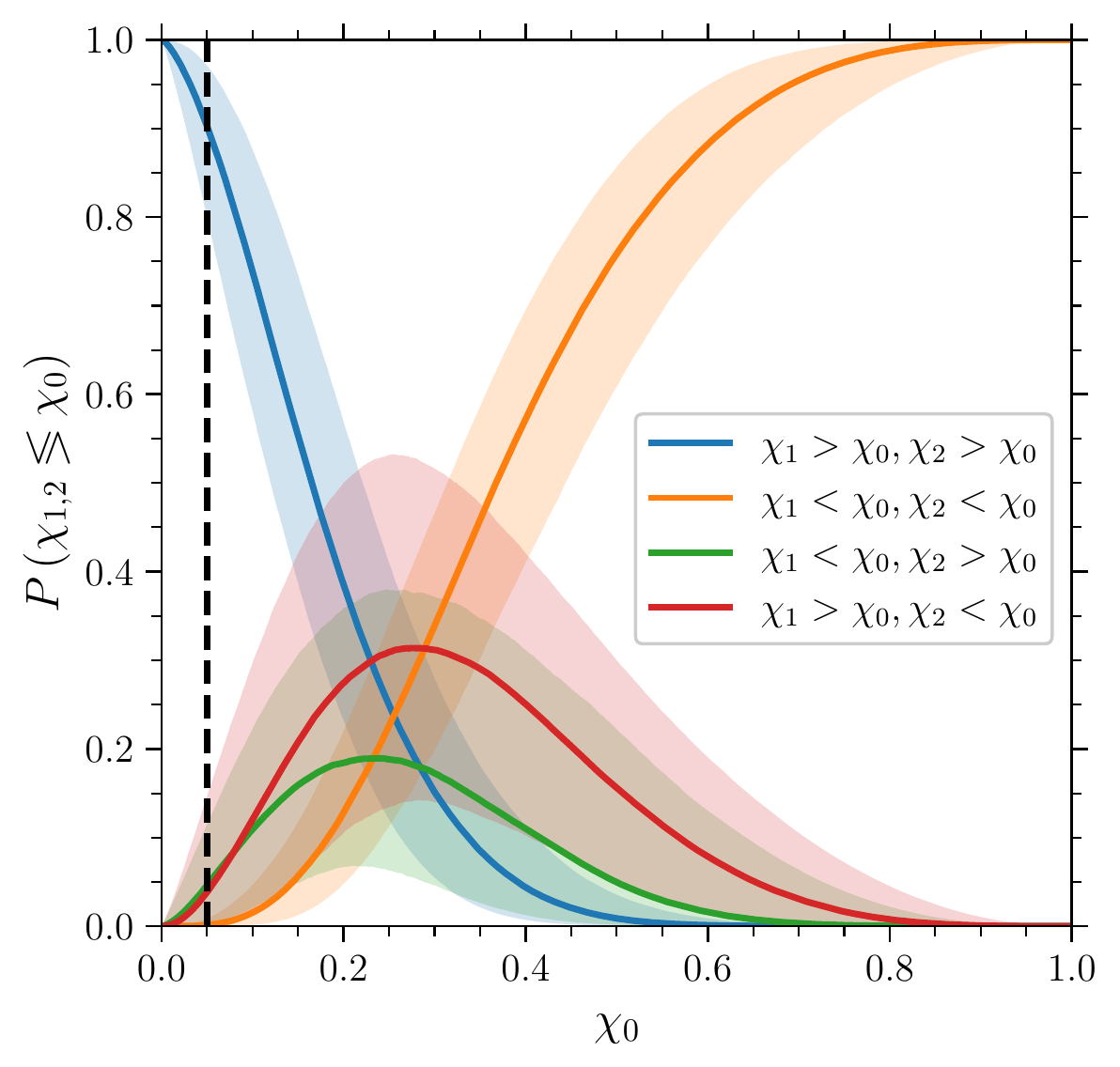}
\caption{The joint probabilities that the spins of the more and less massive BHs, $\chi_1$ and $\chi_2$ respectively, lie above or below a given threshold, $\chi_0$, in the \textsc{nonidentical} model. The blue (orange) curve represents the case in which both components are above (below) threshold. For the red (green) curve, the primary BH spin is above (below) the threshold but the secondary is below (above) it, corresponding to the isolated binary evolution scenario in which MRR has (not) occurred. The solid lines give the median probabilities over the hyperparameter uncertainties as a function of $\chi_0$, while the shaded bands enclose the 90\% credible regions. The vertical dashed black line indicates the fiducial threshold value of $\chi_0=0.05$.}
\label{fig: quadrants}
\end{figure}

In Fig.~\ref{fig: quadrants} we plot the measurements of these probabilities as a function of the spin threshold $\chi_0$; of course, when $\chi_0=1$ ($\chi_0=0$) all sources are classified as having two (non-) negligible spins. There is no support in the population for binaries with \emph{both} spins above $\approx0.6$, while $P(\chi_1,\chi_2<\chi_0|\vlambda)$ dominates for $\chi_0\gtrsim0.8$ since the component spin distributions taper off above this value (see Appendix~\ref{app: spin posteriors}). The fraction of sources that are placed in the MRR ($\chi_1>\chi_0,\chi_2<\chi_0$) and standard ($\chi_1<\chi_0,\chi_2>\chi_0$) scenarios peak at $\chi_0\approx0.3$ and $\chi_0\approx0.25$, respectively. The location of these peaks is determined by the locations of the peaks in the component distributions (see Appendix~\ref{app: spin posteriors}) and the former is slightly higher due to the slight preference for primaries with larger spins.

\subsection{Zero-spin peaks}
\label{sec: results zero-spin peaks}

The above results are inferred using the \textsc{nonidentical} spin model, which does not account for a subpopulation of sources with small spins. Instead of relying on the ad hoc small-spin threshold, the \textsc{nonidentical + zeros} model explicitly allows for fractions $f_i$ ($i=1,2$) occupying a peak at zero spin magnitudes; we find both fractions peak at zero with one-sided 99\% credible bounds $f_1<0.46$ and $f_2<0.36$ (or medians and symmetric 90\% intervals of $f_1 = 0.13_{-0.12}^{+0.22}$ and $f_2 = 0.08_{-0.07}^{+0.18}$) with no posterior support above 0.6 (see Appendix~\ref{app: spin posteriors} for posteriors of the occupation fractions and widths of these peaks). These are consistent with the \textsc{nonidentical} measurements of $P(\chi_i<0.05|\vlambda)$ (recall that the prior bounds on the widths of the zero-spin peaks are $\omega_i<0.05$). For comparison, we compute $P(\chi_1>\chi_2|\vlambda) = 0.56_{-0.21}^{+0.19}$ with $P(\chi_1>\chi_2|\vlambda) > 0.5$ at 68\% confidence for the \textsc{nonidentical + zeros} model, consistent with the \textsc{nonidentical} model and only slightly lowered due to the additional but small contribution from the zero peaks.

As similarly described in Sec.~\ref{sec: results non-identical spins}, the independence of the dimensionless spins means that the fraction of sources with both BHs in the beta distribution component is $f_\beta=(1-f_1)(1-f_2)$ while that for both in the small-spin peaks is $f_0= f_1f_2$. The fraction of binaries in which the primary BH has small spin and the secondary is spinning -- i.e., the standard isolated binary evolution channel -- is $\fS = f_1(1-f_2)$, while the converse -- the MRR scenario -- is $\fR = (1-f_1)f_2$. The spin tilt distributions separate the isolated and dynamical evolution channels with a fraction $\zeta$ of sources in a preferentially aligned component, as predicted for binaries formed in the field, while the rest are isotropic. Therefore, from the total population we can identify the fraction of sources forming in the isolated channel that are placed in each subchannel $X \in \{\beta,0,\mathrm{S},\mathrm{R}\}$ as $\zeta f_X$ (where $\sum_Xf_X = 1$).

\begin{figure}
\includegraphics[width=\columnwidth]{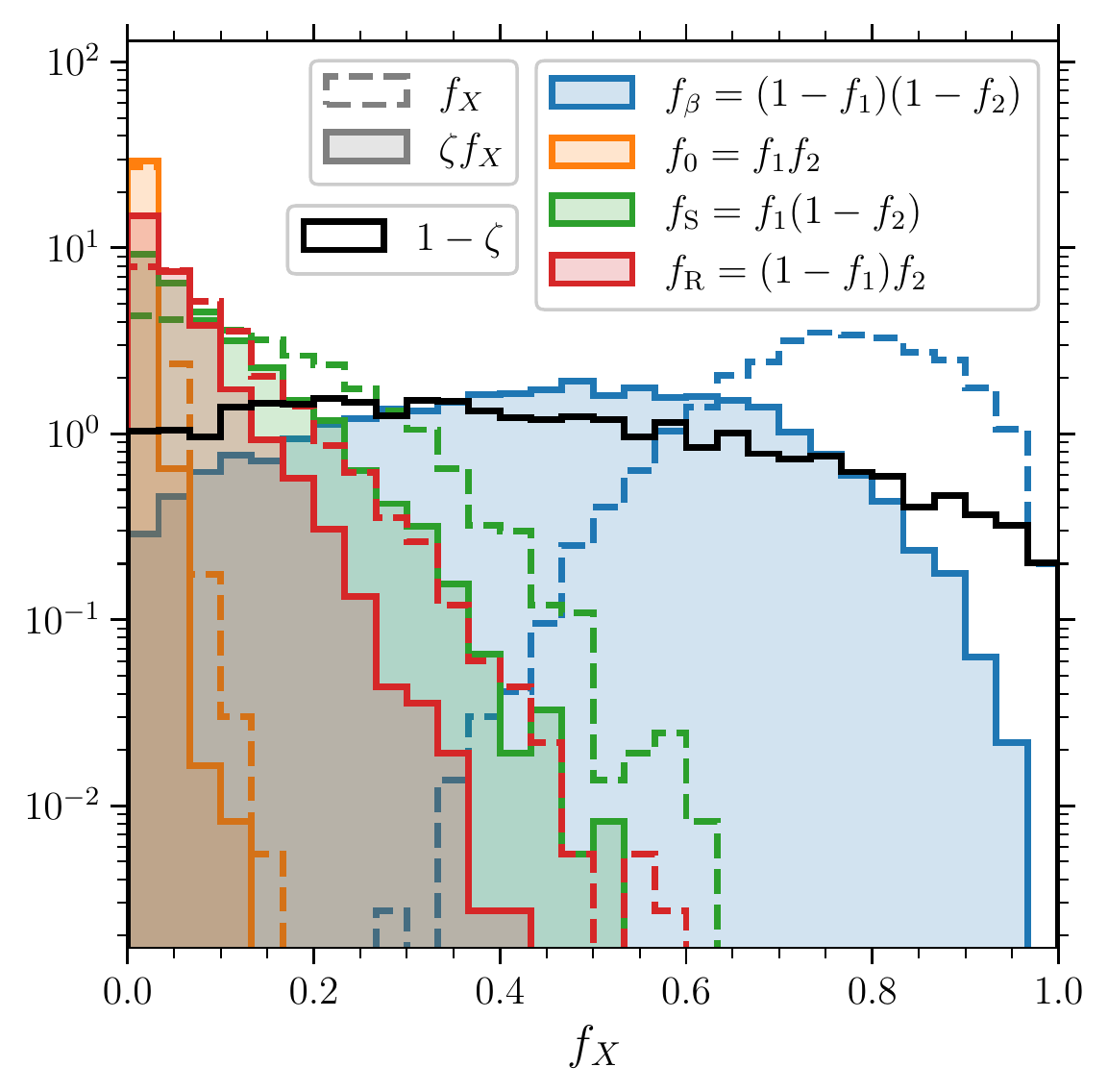}
\caption{Posteriors on the intrinsic fractions of binary BHs in the \textsc{nonidentical + zeros} model with different spin combinations. In blue (orange) is the fraction with two (non-) spinning BHs, $f_\beta$ ($f_0$). In red (green) is the fraction of sources, $\fR$ ($\fS$), in which the more (less) massive BH is spinning while the other is non-spinning and thus interpreted as an MRR (standard) binary. The dashed lines indicate the total fractions within the astrophysical population, $f_X$ ($X\in\{\beta,0,\mathrm{S},\mathrm{R}\}$, $\sum_Xf_X=1$), while the solid filled histograms represent the subset which have preferentially small spin tilts, $\zeta f_X$. The solid black line gives the posterior measurement on the mixing fraction of isotropic spin tilts, $1-\zeta$.}
\label{fig: fractions}
\end{figure}

We present the measurements of these branching fractions from our \textsc{nonidentical + zeros} population analysis in Fig.~\ref{fig: fractions}, including both the total and isolated fractions, $f_X$ and $\zeta f_X$, as well as the fraction of nonisolated sources, $1-\zeta$. Binaries with both BHs placed in the beta spin distributions make up the majority of the population with $f_\beta=0.77_{-0.20}^{+0.16}$. The fractions in which either or both BH spins are negligible all peak at zero: the total standard and MRR fractions are $\fS<0.43$ and $\fR<0.32$ ($\fS=0.12_{-0.11}^{+0.21}$ and $\fR=0.06_{-0.06}^{+0.17}$), respectively, while $f_0<0.06$. The fraction of the total population within the isolated formation channel is not well measured at $\zeta=0.62_{-0.48}^{+0.33}$ (the posterior for $\zeta$ is the same in both the \textsc{nonidentical} and \textsc{nonidentical + zeros} models). Since by definition $0\leq\zeta\leq1$, the isolated fractions are of course lowered, i.e., $\zeta f_X \leq f_X$. In short, at 99\% confidence we measure $\zeta\fR<0.23$, $\zeta\fS<0.33$, and $\zeta f_0<0.04$.

\section{Discussion}
\label{sec: discussion}

\subsection{Comparison to observations and predictions}
\label{sec: comparison}

Altogether, our results (given the current GW catalogue) point to a lack of evidence for large subpopulations of merging stellar-mass binary BHs with negligible spins, or an inability to measure a sharp feature such as this in our models. This is potentially in disagreement both with previous measurements from GW data~\citep{2021ApJ...921L..15G} and predictions of isolated binary evolution~\citep{2012ApJ...759...52D,2013PhRvD..87j4028G,2018A&A...616A..28Q,2019ApJ...881L...1F,Broekgaarden_2022,2022ApJ...933...86Z}.

\cite{2021ApJ...921L..15G} find that $54_{-25}^{+21}\%$ of BHs have negligible spin,\footnote{This value is taken from their updated \href{https://arxiv.org/abs/2109.02424v3}{erratum}.} whereas we find in Sec.~\ref{sec: results zero-spin peaks} that $<46\%$ (36\%) of primary (secondary) BHs have negligible spin with fractions peaking at zero, $23_{-16}^{+20}\%$ of binaries have at least one BH with negligible spin, and $\ssim1\%$ have both BHs with negligible spins. We model negligible spins with a truncated Gaussian distribution of non-zero width centred at zero, whereas their zero-spin peak is a delta distribution at exactly $\chi_{1,2}=0$ resulting from parameter estimation runs performed with non-spinning priors.  They also analysed the binary BH mergers observed up to the end of the first half of O3, whereas the catalogue analysed here contains all binary BHs through the end of O3. To perform the equivalent O3 analysis allowing for non-identical spin distributions would require parameter estimation runs for each event with non-spinning priors placed on either or both BHs, resulting in three additional stochastic-sampling runs per event (and per waveform); such analyses are beyond the scope of our study. However, our results are robust with respect to the choice of population model; we measure consistent fractions of BHs with negligible spins in both the \textsc{nonidentical} and \textsc{nonidentical + zeros} models.

\cite{2022ApJ...937L..13C} recently presented a detailed study on the intrinsic distribution of binary BH spins to address the issue of a potential subpopulation of negligible spins. They focus on the technical details of population modelling and analysis that can lead to erroneous conclusions on such features. They report findings consistent with our own and inconsistent with \cite{2021ApJ...921L..15G}, whether modelling effective or component spin. In particular, as in this work they also find a one-sided posterior measurement peaking at zero for the fraction of BHs with small spin, with an upper limit $<60\%$. In addition to modelling differences, \cite{2022ApJ...937L..13C} find that the Bayes factor between spinning and non-spinning prior hypotheses -- a crucial ingredient in the analysis of \cite{2021ApJ...921L..15G} -- was estimated incorrectly for at least one event. We focus on the astrophysical inferences these results allows us to make, in particular constraining the occurrence of MRR, which unlike in their analysis means crucially we model BH spins as non-identical.

\cite{2022ApJ...933...86Z} use rapid population synthesis to investigate the occurrence of MRR in stellar progenitors of binary BHs. They find that, depending on the efficiency of accretion from the initially more to less massive star, up to 72\% of systems undergo MRR (i.e., have a heavier second-born BH). Similarly, \cite{Broekgaarden_2022} assess both the \emph{detectable} and \emph{intrinsic} MRR fractions in a range of population synthesis models. They find typical intrinsic MRR fractions $>30\%$, lying between 11 and 82\% across their models. These recent results are in general agreement with older predictions which found fractions $\sim10$--50\%~\citep{2012ApJ...759...52D,2018PhRvD..98h4036G}. From GW data we measure the related fraction in the intrinsic population to be $<32\%$ (99\% confidence) based on our definitions and astrophysical assumptions. Consistency between these population synthesis predictions and our measurements therefore requires lower accretion efficiencies between massive stars in order to result in lower MRR rates. However, even assuming a conservatively low accretion efficiency of 0.25, \cite{2022ApJ...933...86Z} find MRR fractions $>18\%$. The low value of our fraction stems from the fact that, even when accounting for the possibility of negligible-spin subpopulations, we infer a small number of negligible-spin BHs in the astrophysical population. This particular MRR scenario requires the secondary BH -- the first-born BH that was initially the more massive star -- to have small spin.

More generally, accounting for the inferred astrophysical distribution of binary BH spins solely with the isolated binary formation channel requires some mechanism to produce sufficiently large spins $\chi_i>0.05$. For example, various scenarios including weak core-envelope coupling, tidal synchronization, and significant accretion can lead to binary BHs with large spins \citep{2021ApJ...921L...2O,2021PhRvD.103f3032S,2022ApJ...930...26S,2022ApJ...933...86Z}. The apparent lack of negligible-spin sources in the intrinsic population may imply that, if isolated formation is the dominant contribution, the assumptions typically made in binary population synthesis are incorrect (e.g., efficient angular momentum transport in massive stars), or other formation channels dominate the GW merger rate. Large spins can be formed in environments where binary BHs interact dynamically, such as dense stellar clusters; hierarchical mergers -- in which binary components are the remnants of previous mergers (see \citealt{2021NatAs...5..749G} for a review) -- lead to high spins with a peak in the distribution around $\chi_i\approx0.7$ (e.g., \citealt{2017ApJ...840L..24F,2017PhRvD..95l4046G,2021ApJ...914L..18D,2021CQGra..38d5012G}).

Nevertheless, this is away from the measured peak in the distributions of dimensionless spins at $\chi_i\approx0.2$--0.3, which is emerging as a solid outcome from several population fits (see \citealt{2021arXiv211103634T,2021ApJ...913L...7A,2021PhRvL.126q1103B,2022PhRvD.105b4076M}, as well as our own Appendix~\ref{app: spin posteriors}). To the best of our knowledge this feature has so far not been identified as a preferred outcome in current stellar models, and therefore requires further attention from the perspective of progenitor population modelling. We note that, though there is no evidence for large subpopulations of binary BHs with negligible spins, this does not mean that individual non-spinning sources do not exist. Indeed, in our \textsc{nonidentical + zeros} model there is posterior support for primary and secondary BHs with spins $\chi_i=0$ at 90\% confidence, a conclusion not allowed by the \textsc{default} and \textsc{nonidentical} models (see Appendix~\ref{app: spin posteriors}). However, given the current catalogue and measurement uncertainties, the two BH spins remain consistent with being drawn from a single continuous distribution.

\subsection{Caveats}
\label{sec: caveats}

Inference results are subject to modelling errors; the implicit conditional in our Bayesian measurements is that of the population model. This is a crucial point, which we have explicitly indicated by reporting the conditional dependence on $\vlambda$ in all probabilities of Sec.~\ref{sec: results}. Indeed, misspecification can lead to biased results~\citep{2022PASA...39...25R} and we do not pretend that the simple parametric models used here fully describe the astrophysical distribution of merging stellar-mass binary BHs.

In particular, the masses, spins, and redshifts are assumed to be independent of each other when in reality they may be correlated~\citep{2021ApJ...922L...5C,2021ApJ...912...98F,2022A&A...665A..59B,2022ApJ...932L..19B}. While the distribution of spin tilts is taken to be a mixture of two contributions, mimicking isolated and dynamical formation, it is likely that multiple channels contribute to the merger rate with distinct features across all source parameters~\citep{2021PhRvD.103h3021W,2021ApJ...910..152Z}. On the other hand, observables such as spin tilts may not be such clear discriminators as previously assumed. For binary BHs formed in isolation, core collapse natal kicks~\citep{2000ApJ...541..319K,2021PhRvD.103f3032S} and mass transfer episodes~\citep{2021PhRvD.103f3007S} may result in significant misalignments and spin precession. Stellar binary processes such as common envelope episodes may also occur in dynamical environments~\citep{2019MNRAS.486.3942K,2020MNRAS.498..495D}, leading to both aligned and misaligned spins~\citep{2021MNRAS.504..910T}. In these cases, the fraction $\zeta$ of sources with preferentially aligned spins employed in Sec.~\ref{sec: results zero-spin peaks} is not a robust discriminator of isolated binary formation.

We chose to model the spin magnitudes independently of the spin orientations. In reality, the spin magnitude distributions may be different even within the two channels that divide the spin orientations, with e.g., the small-spin subpopulation being relevant only for binaries formed in isolation and therefore having preferentially aligned spins. We do not allow for correlations between the spin magnitudes and the mixing fraction $\zeta$, and more complex models would include correlations between these parameters motivated by the astrophysical settings in which stellar-mass binary BH formation occurs. However, given the statistical uncertainties in the current GW catalogue and systematic uncertainties in models of binary BH formation, there is a limit to the available constraining power with a given number of population model parameters. Our \textsc{nonidentical} and \textsc{nonidentical + zeros} models can be considered as minimal extensions to state-of-the-art analyses that may already be demonstrating limited constraining power given the number of model parameters, as seen in the broad measurement of the mixing fraction $\zeta$.

We assume the key MRR signature in GW observations to be spinning primary BHs that formed second from the collapse of the initially less massive stellar progenitor and non-spinning secondary BHs forming first from the initially more massive star. As discussed, several mechanisms can influence the resulting BH spin and therefore the fraction of binaries in the intrinsic population with this spin configuration may not be a clean indicator of the fraction of MRR systems. Contamination can result both from binaries that do not experience MRR but attain our assumed spin configuration and from systems that do undergo MRR but end up merging with different spins.

It is important to stress that the question we are asking the GW data here is, in a nutshell, \emph{``which BH spins the most?''}. The answer is then interpreted as the relative fraction of GW sources that undergo MRR by relying on a (hopefully broad, but admittedly specified) set of astrophysical expectations, thus translating the targeted question into \emph{``which BH formed first?''}.

\section{Conclusions}
\label{sec: conclusions}

Motivated by the possibility of an MRR occurring during the evolution of isolated massive binary stars~\citep{Broekgaarden_2022,2022ApJ...933...86Z}, in which the initially less (more) massive star becomes the (non-) spinning primary (secondary) BH in the binary, we performed population analyses of the confident binary BH mergers detected in GWs by LIGO/Virgo~\citep{2019PhRvX...9c1040A,2021arXiv210801045T,2021arXiv211103606T,2021PhRvX..11b1053A}. Crucially, we used spin models that allow for non-identical distributions between primary and secondary BH spins, as well as subpopulations with negligible spins~\citep{2021ApJ...921L..15G}.

In contrast to \cite{2021ApJ...921L..15G}, we did not find evidence for large subpopulations of negligible-spin BHs. Instead, with current GW data the BH spins are consistent with being drawn from identical, continuous distributions with nonvanishing support at small spins. We measure the occupation fractions of peaks at zero spins to be $<46\%$ and $<36\%$ for primary and secondary BHs, respectively. An independent, complementary study that confirms our findings has been recently presented by \cite{2022ApJ...937L..13C}.

There is vanishing posterior support for binaries with \emph{both} dimensionless spins $>0.6$. We measure the intrinsic fraction of sources in which the heavier BH has significant spin while the lighter does not to be $<32\%$ and peak at zero, and reinterpret this as the fraction of binaries whose progenitors undergo MRR. A small proportion of MRR sources is also in tension with recent predictions from isolated binary evolution~\citep{Broekgaarden_2022,2022ApJ...933...86Z}, implying either that such sources do not contribute to the GW merger rate or that the standard assumptions of isolated binary evolution we made do not hold.

On the other hand, the previously reported overdensity in the dimensionless spin distributions around 0.2--0.3 at present appears to be a robust feature even when accounting for non-identical spins and small-spin subpopulations. A clear explanation for this feature from common formation channels remains to be found.

\section*{Acknowledgements}

We thank Colm Talbot, Jacob Golomb, Simona Miller, Katerina Chatziioannou, Tom Callister, Will Farr, and Monica Colpi for helpful discussions. MM, DG, and NS were supported by European Union's H2020 ERC Starting Grant No. 945155-GWmining, Cariplo Foundation Grant No. 2021-0555, and Leverhulme Trust Grant No. RPG-2019-350. MM and FB acknowledge support from H2020 project HPC-EUROPA3 (INFRAIA-2016-1-730897). Computational work was performed on the BlueBear cluster at the University of Birmingham, and at CINECA with allocations through INFN, Bicocca, and ISCRA Type-B project HP10BEQ9JB.

\section*{Data Availability}
The data underlying this article will be shared on reasonable request to the corresponding author.

\bibliographystyle{mnras_tex_edited}
\bibliography{reversedmassratio}

\appendix

\section{Spin posteriors}
\label{app: spin posteriors}

\begin{figure*}
\includegraphics[width=\columnwidth]{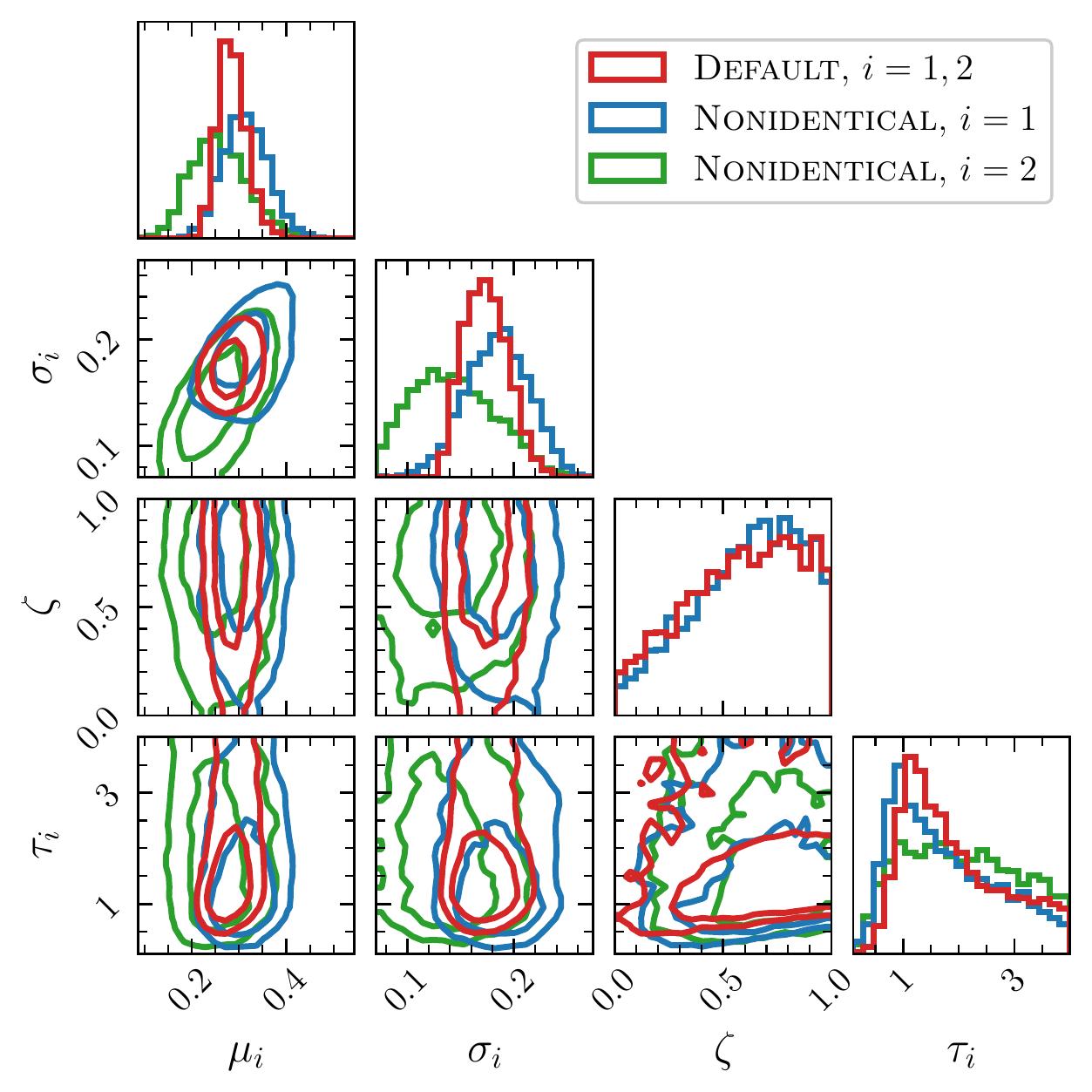}
\includegraphics[width=\columnwidth]{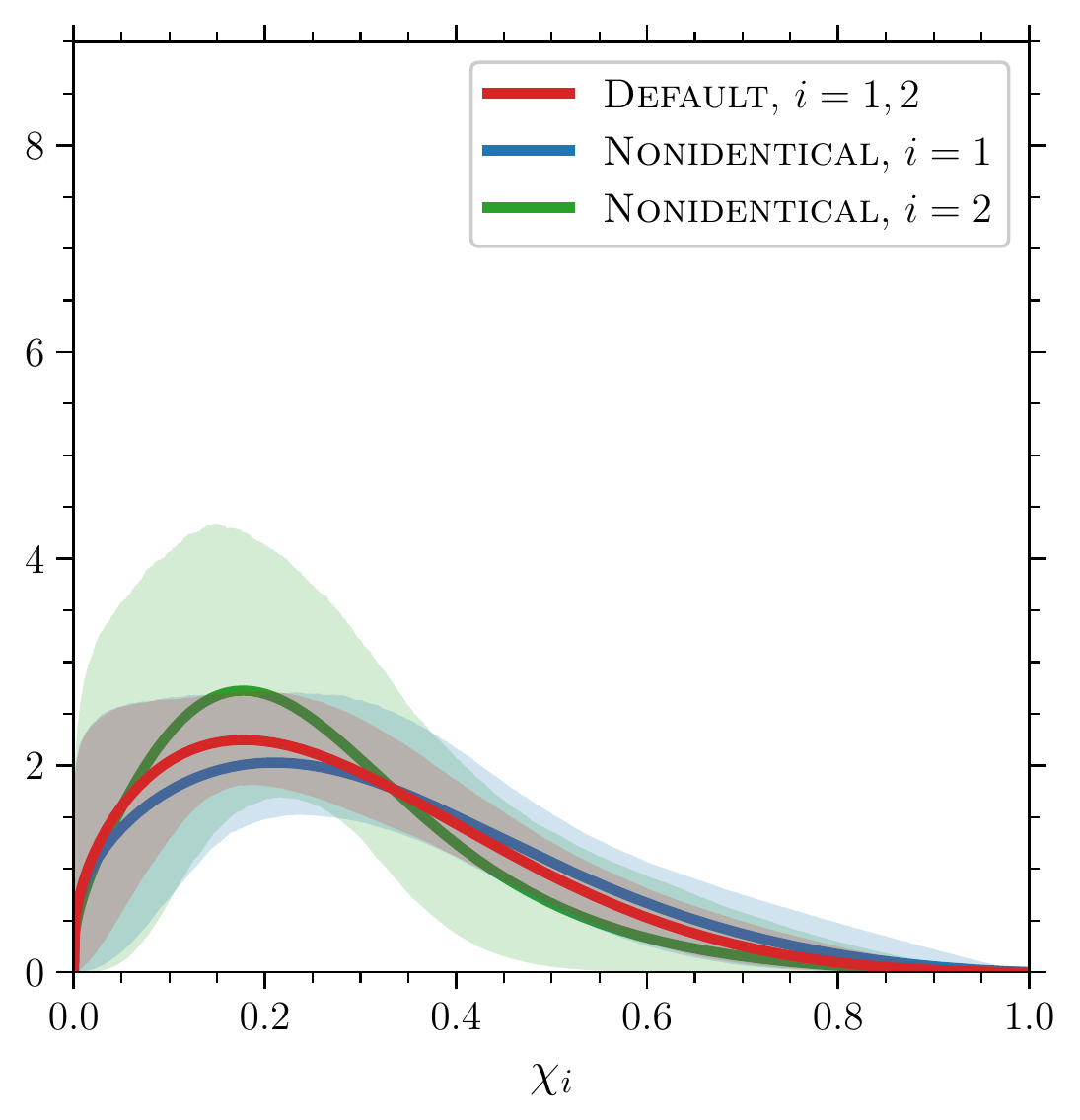}
\caption{The posterior distributions for the spin hyperparameters (left) and the spin magnitudes themselves (right) for the \textsc{nonidentical} model. The primary, $i=1$, distributions (secondary, $i=2$) are plotted in blue (green), and the \textsc{default} spin model is shown in red for comparison. Left: The parameters of the \textsc{nonidentical} model are the means $\mu_i$ and standard deviations $\sigma_i$ of the beta distribution in spin magnitudes, the mixing fraction $\zeta$ between aligned and isotropic spin tilt subpopulations (which is the same for primaries and secondaries), and the widths $\tau_i$ of said aligned components. For the \textsc{default} model the primary and secondary components are identical ($i=1,2$). Right: The solid lines represent the mean distributions, i.e., the posterior population distributions, and the shaded regions correspond to the symmetric 90\% credible regions.}
\label{fig: non-identical posterior}
\end{figure*}

\begin{figure*}
\includegraphics[width=\columnwidth]{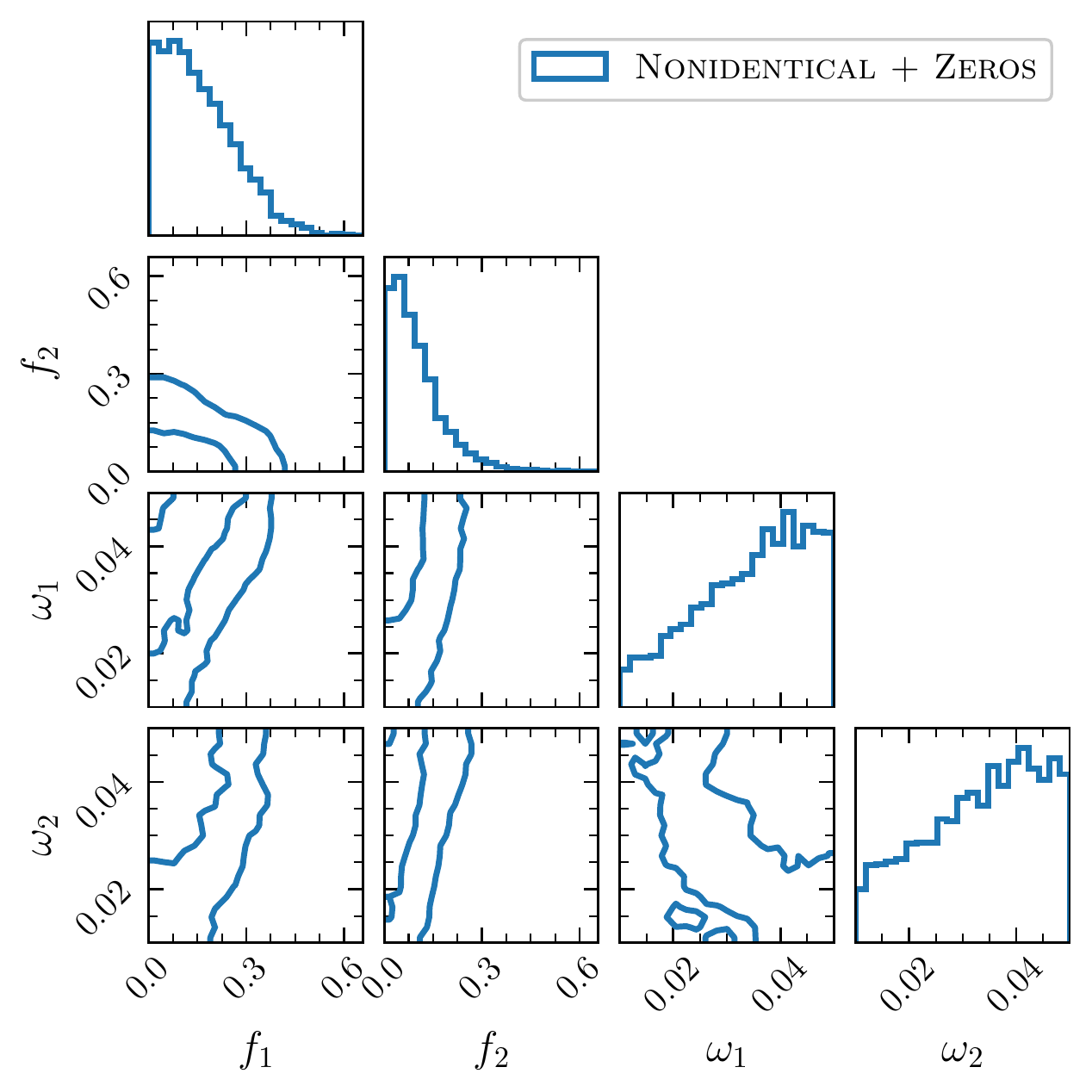}
\includegraphics[width=\columnwidth]{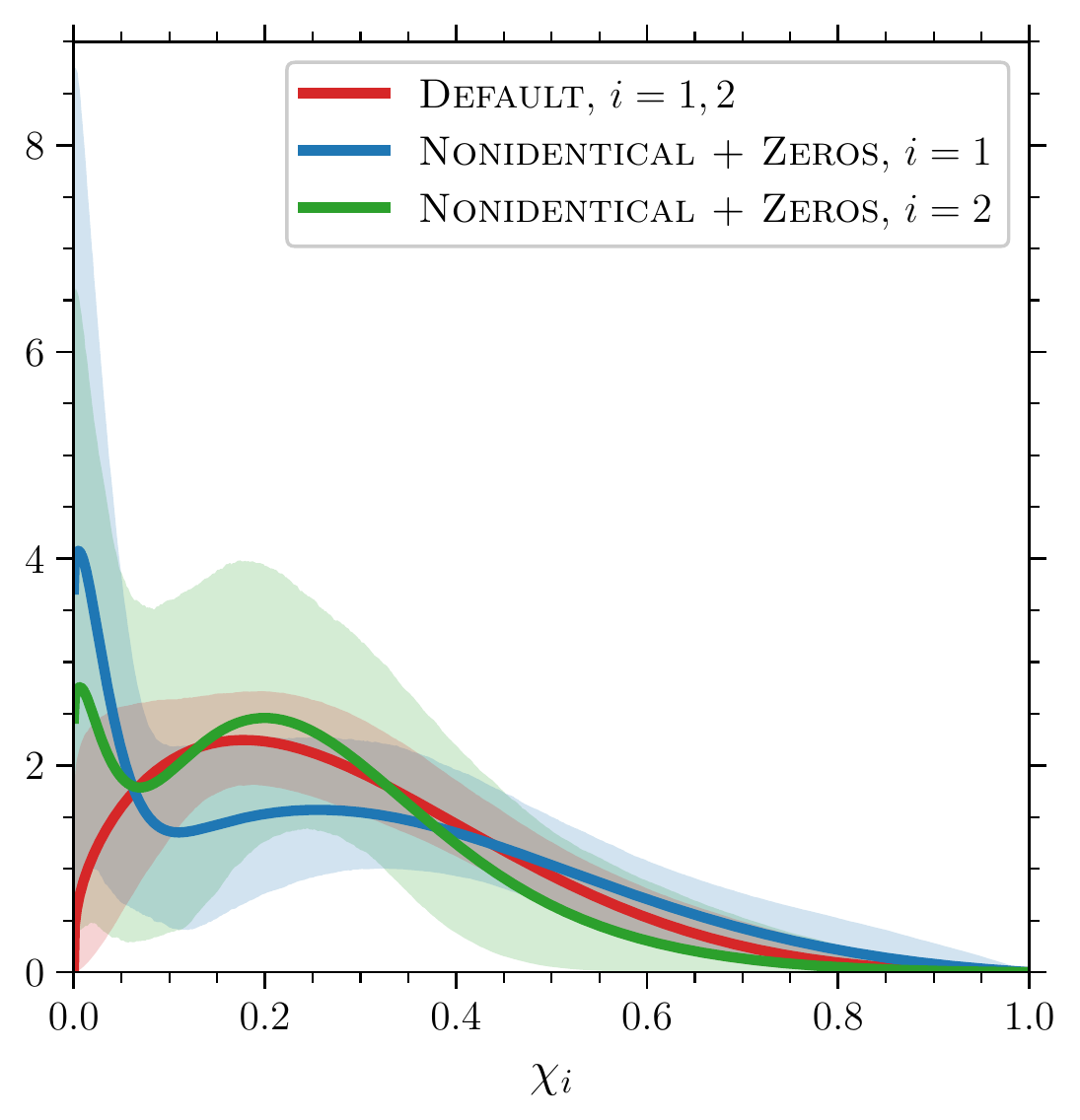}
\caption{Left: The posterior distributions for the hyperparameters of the \textsc{nonidentical + zeros} model characterizing the peaks at zero spins. The occupation fractions of these peaks are $f_i$ and the standard deviations are $\omega_i$ ($i=1,2$). The posteriors of the beta distribution parameters are identical to the \textsc{nonidentical} case and thus not displayed.
Right: The posterior distributions for the spin magnitudes determined by the \textsc{nonidentical + zeros} model. The primary (secondary) spin is plotted in blue (green), and the \textsc{default} model is displayed in red for comparison. The solid lines represent the mean distributions, i.e., the posterior population distributions, and the shaded regions correspond to the symmetric 90\% credible regions.}
\label{fig: non-identicalzero posterior}
\end{figure*}

Here, we present the posterior distributions for the hyperparameters of our models and of the spin magnitudes themselves, including the \textsc{default} spin model for comparison.

Figure~\ref{fig: non-identical posterior} presents the results for the \textsc{nonidentical} model. The parametric forms are the same as in the \textsc{default} model, the only difference being the two BH components are no longer identical. In general this results in larger measurement errors, particular for the secondary spin as seen in the right-hand panel of Fig.~\ref{fig: non-identical posterior}, which is unsurprising since the spin of the less massive component is more difficult to measure in parameter estimation~\citep{2021arXiv211103606T}. We infer secondary BH spin magnitudes with lower means $\mu_2$ and widths $\sigma_2$, resulting in secondary spins which are on average lower and peak more narrowly than that of the primaries. The primary hyperparameters are consistent with the \textsc{default} posteriors, albeit with larger errors. The component spin magnitudes remain consistent with being identical within the displayed 90\% credible regions and feature a peak around $\chi_i\approx0.2$--$0.3$. Given the beta distribution parametrization, the probability densities are consistent with zero at $\chi_i=0$. The mixing fraction $\zeta$ between isotropic and preferentially aligned spin tilts is identical for primary and secondary BHs by assumption -- hence the overlapping of the $i=1,2$ curves in the corresponding one-dimensional histogram -- and is unchanged with respect to the original \textsc{default} analysis. The Bayes factor over the \textsc{default} model is $\log_{10}\mathcal{B}=-0.18$.

In Fig.~\ref{fig: non-identicalzero posterior}, we presents the results for the \textsc{nonidentical + zeros} model that, in addition to the parameters of the \textsc{nonidentical} model, includes Gaussian peaks in each component spin magnitude distribution centred on zero and with standard deviations $\omega_i$ ($i=1,2$) to model a subpopulation of BHs with negligible spin~\citep{2021ApJ...921L..15G}. The posteriors on the parameters of the beta spin components are unchanged with respect to the \textsc{nonidentical} analysis, however, and are hence not plotted; one can simply refer to Fig.~\ref{fig: non-identical posterior} instead. The Bayes factor over the \textsc{default} model is $\log_{10}\mathcal{B}=-0.05$.

The fractions $f_i$ of both primary and secondary BHs with negligible spins are one sided and peak at zero, though the measurement errors are large and extend up to $f_i\lesssim0.6$. The priors on the zero-spin peak widths $\omega_i$ are uniform in $[0.01, 0.05]$ and we find the posteriors feature more support at the upper boundaries. We find the same result when widening the prior domain to $[0.01, 0.1]$, inevitably leading to larger inferred values of $f_i$ but which still peak at zero. The same behaviour is observed in analyses which do not impose a prior cut on the population likelihood determined by the effective sample sizes of Monte Carlo integrals~\citep{2022ApJ...937L..13C}, as used in this work, implying the inferred posteriors are robust. In contrast to the findings of \cite{2021ApJ...921L..15G}, these results altogether point to a lack of evidence for a subpopulation of small spins, or alternatively an inability to measure such a feature in the spin distribution. On the other hand, the peak at $\chi_2\approx0.2$--0.3 remains apparent in the right-hand panel of Fig.~\ref{fig: non-identicalzero posterior}. While allowing for the possibility of a subpopulation of BHs with small spins results in a visible spike in density at $\chi_i=0$, the probability there is still low due to the small volume of occupied parameter space ($\chi_i<0.05$). Unlike the \textsc{default} and \textsc{nonidentical} models, the mixture parametrization of the \textsc{nonidentical + zeros} model allows the $\chi_i$  distributions to be constrained away from zero at $\chi_i=0$ within the 90\% confidence interval. Within this uncertainty, both primary and secondary BH spins are still consistent with a single continuous distribution, however.

\end{document}